\documentclass[preprint,aps]{revtex4}

\newcommand{\cC}{\ensuremath{\mathcal{C}}}
\newcommand{\cP}{\ensuremath{\mathcal{P}}}

\newcommand{\cT}{\ensuremath{\mathcal{T}}}
\newcommand{\cQ}{\ensuremath{\mathcal{Q}}}

\begin{document}

\title{$\mathcal{P}\mathcal{T}$ symmetry and necessary and
sufficient conditions for the reality of energy eigenvalues}

\author{Carl~M.~Bender${}^1$ and Philip~D.~Mannheim${}^2$}

\affiliation{${}^1$Physics Department\\ Washington University\\ St.~Louis, MO
63130, USA\\ {\tt electronic address: cmb@wustl.edu}\\ \\
${}^2$Department of Physics\\ University of Connecticut\\ Storrs, CT 06269, USA
\\{\tt electronic address: philip.mannheim@uconn.edu}}

\date{8 February 2009}

\begin{abstract}
Despite its common use in quantum theory, the mathematical requirement of Dirac
Hermiticity of a Hamiltonian is sufficient to guarantee the reality of energy
eigenvalues but not necessary. By establishing three theorems, this paper gives
physical conditions that are both necessary and sufficient. First, it is shown
that if the secular equation is real, the Hamiltonian is necessarily $\cP\cT$
symmetric. Second, if a linear operator $\cC$ that obeys the two equations
$[\cC,H]=0$ and $\cC^2=1$ is introduced, then the energy eigenvalues of a $\cP
\cT$-symmetric Hamiltonian that is diagonalizable are real only if this $\cC$
operator commutes with $\cP\cT$. Third, the energy eigenvalues of $\cP
\cT$-symmetric Hamiltonians having a nondiagonalizable, Jordan-block form are
real. These theorems hold for matrix Hamiltonians of any dimensionality.
\end{abstract}

\maketitle

\section{Introduction}
\label{S1}

In quantum mechanics physical observables must have real eigenvalues. Thus, it
is conventional, and even axiomatic, to associate such observables with
Dirac-Hermitian operators. (A {\it Dirac-Hermitian} operator is invariant under
combined matrix transposition and complex conjugation. The Dirac adjoint of $A$
is written $A^\dag$.) However, while a Dirac-Hermitian operator has real
eigenvalues, there is nothing that prevents a non-Dirac-Hermitian operator from
also having real eigenvalues. Thus, Dirac Hermiticity is sufficient to ensure
the reality of eigenvalues but it is not necessary. In this paper we provide
conditions that are both necessary and sufficient to guarantee real eigenvalues.

Interest in non-Dirac-Hermitian operators having real eigenvalues was triggered
by the discovery that the energy eigenvalues of Hamiltonians such as $H=p^2+i
x^3$ are all real \cite{R1,R2,R2a}. Because the parity operator $\cP$ transforms
$x^3$ into $-x^3$ and the time reversal operator $\cT$ transforms $i$ into $-i$,
the reality of the energy eigenvalues of $H=p^2+ix^3$ was traced in
Ref.~\cite{R2} to the facts that the Hamiltonian is $\cP\cT$ invariant and that
its eigenstates are also eigenstates of the $\cP\cT$ operator. Specifically, for
an energy eigenstate, which satisfies $H|\psi\rangle=E|\psi\rangle$, the
vanishing of the $[H,\cP\cT]$ commutator and the antilinearity of the $\cT$
operator imply that $\cP\cT H|\psi\rangle =E^*\cP\cT|\psi\rangle=H\cP\cT|\psi
\rangle$. Since $\cP\cT|\psi\rangle$ is equal (apart from an irrelevant phase)
to $|\psi \rangle$, one concludes that $E=E^*$. The Hamiltonian $H=p^2+ix^3$
thus emerges as a non-Dirac-Hermitian operator whose eigenvalues are all real.

However, even if $H$ and $\cP\cT$ commute, there is no guarantee that the
eigenvalues of $H$ are real because the eigenstates of $H$ need not be
eigenstates of the $\cP\cT$ operator; $\cP\cT$-symmetric Hamiltonians that are
not Dirac Hermitian may possess complex as well ad realeigenvalues. (The
familiar theorem that one can simultaneously diagonalize any two commuting
operators does not hold when antilinear operators are involved.) Nonetheless,
$\cP\cT$ symmetry is still a powerful condition because it implies that the
secular equation ${\rm det}(H-EI)=0$, which determines the eigenvalues of the
Hamiltonian, is real \cite{R3}.

To establish the reality of the secular equation of a $\cP\cT$-symmetric
Hamiltonian, one can use standard properties of determinants: ${\rm det}(H-EI)=
{\rm det}(\cP\cT H\cT^{-1}\cP^{-1}-EI)={\rm det}(\cT H\cT^{-1}-EI)$. Then,
writing the time-reversal operator as $\cT=LK$ and $\cT^{-1}=KL^{-1}$, where $L$
is a linear operator and $K$ performs complex conjugation, one obtains ${\rm det}
(H-EI)={\rm det}(H^*-EI)$. Thus, $H$ and $H^*$ have the same set of eigenvalues
and $H$ has a real secular equation. Note that this conclusion holds for any
choice of linear operators $\cP$ and $L$. Thus, we can interpret the condition
of $\cP\cT$ symmetry in the generalized sense, where $\cP$ is any linear
operator and $\cT$ is any antilinear operator. These generalized operators
$\cP$ and $\cT$ need not even commute with one another.

There is an even shorter proof that the secular equation is real. The secular
equation is a polynomial in the eigenvalue $E$ of the form $\sum_n a_n E^n=0$.
The Hamiltonian matrix itself solves this equation: $\sum_n a_n H^n=0$. If the
Hamiltonian commutes with {\it any} antilinear operator $\kappa$, $[H,
\kappa]=0$, then $H$ also obeys $\sum_n a_n^* H^n=0$. Thus, the secular equation
is real. Establishing the reality of the secular equation does not require that
$\kappa^2=1$.

The $\cP\cT$ symmetry of a Hamiltonian thus requires that any energy eigenvalue
must either be real or belong to a complex-conjugate pair. If the eigenvalue is
real, the associated eigenstate is also an eigenstate of the $\cP\cT$ operator.
If the energy is a member of a complex conjugate pair, the $\cP\cT$ operator
maps its eigenstate into the eigenstate associated with the complex-conjugate
energy eigenvalue.

{}From the discussion above we see that $\cP\cT$ symmetry alone is not powerful
enough to force the energy eigenvalues of a Hamiltonian to be real. Moreover,
while $\cP\cT$ symmetry implies a real secular equation, it does not follow that
$\cP\cT$ symmetry is the only way to produce a real secular equation; $\cP\cT$
symmetry is sufficient but not necessary to give a real secular equation.

In this paper we prove three theorems: (i) If the secular equation is real, the
Hamiltonian is necessarily $\cP\cT$ symmetric. Thus, $\cP\cT$ symmetry is both
necessary and sufficient for the reality of the secular equation, and it is not
possible for a Hamiltonian that is not $\cP\cT$-symmetric to possess an entirely
real set of energy eigenvalues. (ii) Introduce the $\cC$ operator, which is
linear and obeys $[\cC,H]=0$ and $\cC^2=1$ \cite{R4,R5}. Then, the energy
eigenvalues of a $\cP\cT$-symmetric Hamiltonian that is diagonalizable are real
only if this $\cC$ operator commutes with $\cP\cT$. (iii) The energy eigenvalues
of $\cP\cT$-symmetric Hamiltonians having a nondiagonalizable, Jordan-block form
are real. These theorems hold for Hamiltonians of any dimensionality, but
because a pair of complex-conjugate eigenvalues span a two-dimensional space,
one can derive these results by considering two-dimensional Hamiltonians.

In Sec.~\ref{S2} the properties of two-dimensional $\cP\cT$-symmetric matrices
are discussed. The $\cC$ operator is introduced in Sec.~\ref{S3},
nondiagonalizable Jordan-block matrices are examined in Sec.~\ref{S4}, and the
$\cC\cP=e^\cQ$ operator of $\cP\cT$ quantum mechanics is studied in
Sec.~\ref{S5}.

\section{Two-dimensional Matrices}
\label{S2}

The Pauli matrices form a basis of $2\times2$ matrices, so in a $2\times2$ space
we can take general Hamiltonian, parity, and time-reversal operators to have the
form
\begin{equation}
H=\sigma_0h^0+\mbox{\boldmath $\sigma$}\cdot{\bf h},\qquad\cP=\sigma_0p^0+
\mbox{\boldmath $\sigma$}\cdot{\bf p},\qquad \cT=K\sigma_2(\sigma_0
t^0+\mbox{\boldmath $\sigma$}\cdot{\bf t}).
\label{E1}
\end{equation}
Here the factor $\sigma_2$ has been introduced in $\cT$ because of its
convenient property $\sigma_2 \mbox{\boldmath $\sigma$}\sigma_2=-\mbox{\boldmath
$\sigma$}^*$. Since $h^0=h_{\bf R}^0+ih_{\bf I}^0$ and ${\bf h}={\bf h_R}+i{\bf
h_I}$ are not required to be real, the Hamiltonian $H$ is not necessarily Dirac
Hermitian. For  $\cP$ and $\cT$ to serve as the conventional parity and
time-reversal operators they need to obey $\cP^2=1$, $\cT^2=1$, $[\cP,\cT]=0$,
with $\cP$ being Dirac Hermitian and unitary and $\cT$ being expressible as $K$
times a unitary operator. These conditions can be satisfied if we take $p^0=0$,
${\bf p}\cdot{\bf p}=1$, ${\bf p=p^*}$, $t^0=0$, ${\bf t}\cdot{\bf t}=1$, ${\bf
t=t^*}$, and require that ${\bf p}\cdot{\bf t}=0$. The vectors ${\bf p}$ and
${\bf t}$ are thus real, orthogonal unit vectors.

For the Hamiltonian (\ref{E1}), the secular equation for the energy eigenvalues
$E$ is
\begin{equation}
E^2-2Eh^0+(h^0)^2={\bf h}\cdot{\bf h}={\bf h_R}\cdot{\bf h_R}-{\bf h_I}\cdot{\bf
h_I}+2i{\bf h_R}\cdot{\bf h_I}.
\label{E2}
\end{equation}
Furthermore, since $({\bf \sigma}\cdot{\bf A})(\mbox{\boldmath $\sigma$}\cdot
{\bf B})(\mbox{\boldmath $\sigma$}\cdot{\bf A})=2(\mbox{\boldmath $\sigma$}\cdot
{\bf A})({\bf A}\cdot{\bf B})-(\mbox{\boldmath $\sigma$}\cdot{\bf B})({\bf A}
\cdot{\bf A})$ for any two vectors ${\bf A}$ and ${\bf B}$, we obtain
\begin{equation}
\cP\cT H\cT^{-1}\cP^{-1}-H=-2i\sigma_0h_{\bf I}^0+2\mbox{\boldmath $\sigma$}
\cdot{\bf F}-2i\mbox{\boldmath $\sigma$}\cdot{\bf G},
\label{E3}
\end{equation}
where
\begin{equation}
{\bf F}=({\bf h_R}\cdot{\bf p}){\bf p}+({\bf h_R}\cdot{\bf t}){\bf t}-{\bf h_R},
\qquad{\bf G}=({\bf h_I}\cdot{\bf p}){\bf p}+({\bf h_I}\cdot{\bf t}){\bf t}.
\label{E4}
\end{equation}

If we now assume that the Hamiltonian is $\cP\cT$ invariant, then from
(\ref{E3}) we obtain $h_{\bf I}^0=0$, ${\bf F=0}$, ${\bf G=0}$. Since ${\bf p}
\cdot{\bf t}=0$, the vanishing of ${\bf G}$ implies that both ${\bf h_I}\cdot
{\bf p}$ and ${\bf h_I}\cdot{\bf t}$ vanish. Then, from the vanishing of ${\bf
F}$ we see that ${\bf h_R}\cdot{\bf h_I}=0$. Consequently, the secular equation
(\ref{E2}) is strictly real, and we recover the result of \cite{R3}
for $\cP\cT$-symmetric Hamiltonians.

On the other hand, suppose we start by assuming that the secular equation is
real and set $h_{\bf I}^0=0$ and ${\bf h_R}\cdot{\bf h_I}=0$. The vanishing of
${\bf h_R}\cdot{\bf h_I}$ can be achieved in three ways: ${\bf h_I=0}$, ${\bf
h_R=0}$, ${\bf h_I}$ and  ${\bf h_R}$ orthogonal. If ${\bf h_I=0}$, and thus
${\bf G=0}$, the choice ${\bf p}={\bf h_R}/({\bf h_R}\cdot{\bf h_R})^{1/2}$ then
yields ${\bf h_R}\cdot{\bf t}=0$ and thus ${\bf F=0}$. If ${\bf h_R=0}$, and
thus ${\bf F=0}$, because ${\bf h_I}$ is a 3-vector and ${\bf p}\cdot{\bf t}=0$,
for any ${\bf h_I}$ we can pick an appropriate ${\bf p}$ and ${\bf t}$ so that
${\bf h_I}$, ${\bf p}$, and ${\bf t}$ form an orthogonal triad. This then yields
${\bf G=0}$. If both ${\bf h_R}$ and ${\bf h_I}$ are nonzero, the choice ${\bf
p}={\bf h_R}/({\bf h_R} \cdot{\bf h_R})^{1/2}$ then yields ${\bf h_R}\cdot {\bf
t}=0$ and ${\bf F=0}$. Moreover, this choice for ${\bf p}$ leads to ${\bf h_I}
\cdot{\bf p}=0$ because ${\bf h_R}\cdot{\bf h_I}=0$, and thus to ${\bf G}=({\bf
h_I}\cdot{\bf t}){\bf t}$. With ${\bf p}$ being parallel to ${\bf h_R}$, the
vectors ${\bf t}$ and ${\bf h_I}$ must lie in a plane that is normal to ${\bf
h_R}$. For any choice of ${\bf h_I}$ in that plane we can always find a ${\bf
t}$ that is orthogonal to it, and thus make ${\bf G}$ vanish. Hence, for any
choice of ${\bf h_R}$ and ${\bf h_I}$, there will always be a choice of ${\bf
p}$ and ${\bf t}$ for which $H$ is $\cP\cT$ symmetric. Therefore, $\cP\cT$
invariance of $H$ is both necessary and sufficient to ensure the reality of all
the coefficients in its secular eigenvalue equation.

We have shown that the reality of the secular equation implies the $\cP\cT$
symmetry of the Hamiltonian in the two-dimensional case. To extend this result
to higher-dimensional matrices we need only assume that the Hamiltonian can be
diagonalized. (Remember that $\cP\cT$-symmetric Hamiltonians have real secular
equations regardless of their dimensionality.) In its diagonal form a
Hamiltonian whose secular equation is real consists of a set of real eigenvalues
and a set of complex-conjugate pairs of eigenvalues $E^i_{\bf R}\pm iE^i_{\bf
I}$ on the diagonal. For the real eigenvalues $\cP\cT$ symmetry is made manifest
by simply taking $\cP$ to be diagonal and $\cT$ to be $K$ in that sector of the
Hamiltonian. In each two-dimensional block of complex-conjugate eigenvalues the
Hamiltonian behaves as $H^i=E^i_{\bf R}\sigma_0+i E_{\bf I}^i\sigma_3$ and it is
thus $\cP\cT$ invariant block-by-block under $\cP=\sigma_1$ and $\cT=K$
[equivalent to taking ${\bf p}=(1,0,0)$, ${\bf t}=(0,1,0)$ in (\ref{E1})].
Transforming back to the original, undiagonalized form of the Hamiltonian then
yields a Hamiltonian that is symmetric under the transformed $\cP\cT$ operator.
The theorem on the secular equation thus holds for diagonalizable Hamiltonians
of arbitrary dimensionality. To determine the conditions for which the roots of
the now real secular equation are all real, we introduce the $\cC$ operator.

\section{The $\cC$ operator}
\label{S3}

In previous studies a linear operator $\cC$ was introduced that satisfied the
three conditions
\begin{equation}
[\cC,H]=0,\qquad\cC^2=1,\qquad[\cC,\cP\cT]=0.
\label{E5}
\end{equation}
The $\cC$ operator was then used to construct a Hilbert-space inner product for
$\cP\cT$-symmetric Hamiltonians whose energy eigenvalues were all real. The
associated $\cC\cP\cT$ norm $\langle\psi|\cC\cP\cT|\psi\rangle$ was positive
definite. For our purposes here we do not impose the third condition in
(\ref{E5}) and instead introduce a generalized linear $\cC$ operator that obeys
just the two conditions
\begin{equation}
[\cC,H]=0,\qquad\cC^2=1.
\label{E6}
\end{equation}
Unlike the $\cP$ operator, the $\cC$ operator is not required to be Dirac
Hermitian.

To construct this generalized operator $\cC$ in the two-dimensional case we set
$\cC=\sigma_0c^0+\mbox{\boldmath $\sigma$}\cdot{\bf c}$, and find that we need
to impose the conditions
\begin{equation}
c^0=0,\qquad {\bf c}\cdot{\bf c}=1,\qquad({\bf c}\cdot{\bf h}){\bf c}-{\bf h}=0.
\label{E8}
\end{equation}
(We exclude the trivial solution $\cC=1$.) Equation (\ref{E8}) shows that ${\bf
c}={\bf h}/({\bf h}\cdot{\bf h})^{1/2}$ unless ${\bf h}\cdot{\bf h}$ happens to
vanish. Because $[\cC,H]=0$ our analysis extends to diagonalizable Hamiltonians
of arbitrary dimensionality and establishes that apart from the special case
${\bf h}\cdot{\bf h}=0$, a nontrivial $\cC$ operator satisfying $[\cC,H]=0$,
$\cC^2=1$ exists \cite{R6}.

For the special case in which ${\bf h}\cdot{\bf h}$ vanishes, the conditions
required by (\ref{E8}) cannot be realized because a nonzero ${\bf h}$ would
require a nonzero ${\bf c}\cdot{\bf h}$, while a vanishing ${\bf h}\cdot{\bf h}$
would require a vanishing $({\bf c}\cdot{\bf h})^2$. Thus, when ${\bf h}\cdot{
\bf h}=0$, $\cC$ is undefined. (We defer further discussion of the ${\bf h}\cdot
{\bf h}=0$ case to Sec.~\ref{S4}. Unlike $\cC$, both $\cP$ and $\cT$ remain
well-defined at ${\bf h}\cdot{\bf h}=0$.)

To explore the implications of (\ref{E8}) it is convenient to set ${\bf c}\cdot{
\bf h}=X+iY$, where
\begin{equation}
{\bf c_R}\cdot{\bf h_R}-{\bf c_I}\cdot{\bf h_I}=X,\qquad{\bf c_R}\cdot{\bf h_I}+
{\bf c_I}\cdot{\bf h_R}=Y.
\label{E9}
\end{equation}
Since the condition $({\bf c}\cdot {\bf h}){\bf c}-{\bf h}=0$ forbids the
vanishing of ${\bf c}\cdot {\bf h}$ once ${\bf h}$ is nonzero, at least one of
the parameters $X$ and $Y$ must be nonzero. {}From (\ref{E8}) we obtain
\begin{equation}
{\bf c_R}\cdot{\bf c_R}-{\bf c_I}\cdot{\bf c_I}=1,\qquad {\bf c_R}\cdot{\bf c_I}
=0,\qquad X{\bf c_R}-Y{\bf c_I}={\bf h_R},\qquad Y{\bf c_R}+X{\bf c_I}={\bf
h_I},
\label{E10}
\end{equation}
and because $X^2+Y^2$ cannot be zero we get ${\bf c}=(X-iY){\bf h}/(X^2+Y^2)$.
{}From (\ref{E10}) we obtain
\begin{equation}
{\bf h_R}\cdot{\bf h_I}=XY,\qquad {\bf h_R}\cdot{\bf h_R}-{\bf h_I}\cdot{\bf h_I
}=X^2-Y^2.
\label{E12}
\end{equation}
When the Hamiltonian is $\cP\cT$ symmetric, ${\bf h_R}\cdot{\bf h_I}$ is
required to vanish, and we see that the $XY$ product also vanishes. Thus, either
$X$ and $Y$, but not both, must vanish. Since the solutions to the eigenvalue
equation (\ref{E2}) have the form $E=h^0\pm({\bf h_R}\cdot{\bf h_R}-{\bf h_I}
\cdot{\bf h_I})^{1/2}$ when ${\bf h_R}\cdot{\bf h_I}=0$, for $\cP\cT$-symmetric
Hamiltonians (where $h_{\bf I}^0=0$) the energy eigenvalues will both be real if
we realize the condition $XY=0$ via $Y=0$, and they will form a
complex-conjugate pair if we set $X=0$. Thus, to complete the proof we must
relate the vanishing of $Y$ or $X$ to the vanishing or nonvanishing of the
$[\cC,\cP\cT]$ commutator.

We evaluate the $[\cC,\cP\cT]$ commutator in the two cases and obtain
\begin{equation}
\cP\cT\cC\cT^{-1}\cP^{-1}-\cC=\frac{2(X+iY)}{X^2+Y^2}\mbox{\boldmath $\sigma$}
\cdot({\bf F}-i{\bf G})+\frac{2iY}{X^2+Y^2}\mbox{\boldmath $\sigma$}\cdot{\bf
h}.
\label{E13}
\end{equation}
Since ${\bf F}$ and ${\bf G}$ vanish when $H$ is $\cP\cT$ symmetric, $\cP\cT\cC
\cT^{-1}\cP^{-1}-\cC$ vanishes only if $Y=0$. We thus establish that when $\cP
\cT$ commutes with $H$ and $\cC$, all energy eigenvalues are real.

It is of interest to evaluate the $\{\cC,\cP\cT\}$ anticommutator as well, and
it has the form
\begin{equation}
\cP\cT\cC\cT^{-1}\cP^{-1}+\cC=\frac{2(X+iY)}{X^2+Y^2}\mbox{\boldmath $\sigma$}
\cdot({\bf F}-i{\bf G})+\frac{2X}{X^2+Y^2}\mbox{\boldmath $\sigma$}\cdot{\bf h}.
\label{E14}
\end{equation}
Thus, when $H$ is $\cP\cT$ symmetric, $\cP\cT\cC\cT^{-1}\cP^{-1}+\cC$ vanishes
only if $X=0$. Therefore, when $X=0$, $\cC$ and $\cP\cT$ anticommute rather
than commute. Hence, when $\cP\cT$ commutes with $H$ but not with $\cC$, the
energy eigenvalues appear in complex-conjugate pairs \cite{R11}.

Finally, since $\cC$ and $H$ commute, our result immediately generalizes to
diagonalizable Hamiltonians of arbitrary dimensionality. Specifically, one first
constructs the appropriate $\cC$ operator in the basis in which both $\cC$ and
$H$ are diagonal, and then one transforms back to the original nondiagonal
basis. Thus, except when the Hamiltonian is not diagonalizable, we have shown
that energy eigenvalues are real only if  $[H,\cP\cT]=0$ and $[\cC,\cP\cT]=0$.
We turn next to the case of nondiagonalizable Hamiltonians.

\section{Nondiagonalizable Hamiltonians}
\label{S4}

Dirac-Hermitian Hamiltonians can always be diagonalized by means of a unitary
transformation, but a Hamiltonian that is not Dirac Hermitian may not be
diagonalizable. Jordan showed that via a sequence of similarity transformations
any $N$-dimensional square matrix can be brought to either a diagonal form or a
triangular Jordan-block form in which all of the elements on one side of the
diagonal are zero. Since the nonzero elements on the other side of the diagonal
of a Jordan-block matrix do not contribute to the secular equation, the $N$
elements on its diagonal are the $N$ eigenvalues of the matrix. Because
Jordan-block matrices cannot be brought to a diagonal form by a similarity
transform, they possess fewer than $N$ eigenvectors (even though they possess a
full set of $N$ eigenvalues). Thus, the eigenvectors of a Jordan-block matrix do
not form a complete basis.

A Jordan-block matrix has fewer solutions to the eigenvector equation $H\psi=E
\psi$ than there are to the determinantal condition ${\rm det}(H-EI)=0$. For
diagonalizable Hamiltonians the eigenvalue equation $H\psi=E\psi$ yields $N$
eigenvalues, and so one does not need to distinguish between eigenvalue
solutions to $H\psi=E\psi$ and eigenvalue solutions to ${\rm det}(H-EI)=0$.
However, for nondiagonalizable Hamiltonians one does need to make a distinction,
and we shall thus refer to $H\psi=E\psi$ as the {\it eigenvector equation} and
to ${\rm det}(H-EI)=0$ as the {\it eigenvalue} or {\it secular equation}.

A typical two-dimensional example of a Jordan-block matrix is
\begin{eqnarray}
M=\pmatrix{a&b\cr 0&a}.
\label{E15}
\end{eqnarray}
This matrix possesses two eigenvalues, both equal to $a$, but has only one
eigenvector because the eigenvector condition
\begin{eqnarray}
\pmatrix{a&b\cr 0&a}\pmatrix{c \cr d}=\pmatrix{ac+bd\cr ad}=\pmatrix{ac\cr ad}
\label{E16}
\end{eqnarray}
only has one solution, namely, that with $d=0$. When the parameter $a$ is real,
(\ref{E15}) provides a simple example of a non-Hermitian matrix whose
eigenvalues are all real.

Jordan-block matrices must have degenerate eigenvalues. Note that
if we replace the matrix $M$ of (\ref{E15}) by the non-Jordan-block matrix
\begin{eqnarray}
M(\epsilon)=\pmatrix{a&b\cr 0&a+\epsilon},
\label{E17}
\end{eqnarray}
we find two eigenvalues $a$ and $a+\epsilon$ and two independent eigenvectors:
\begin{eqnarray}
\psi(a)=\pmatrix{1\cr0},\qquad\psi(a+\epsilon)=\pmatrix{1\cr\epsilon/b}.
\label{E18}
\end{eqnarray}
However, in the limit $\epsilon\to0$ the two eigenvectors merge and one
eigenvector is lost \cite{R7}.

The theorem requiring the secular equation to be real if $[H,\cP\cT]=0$ is not
sensitive to Jordan-block structures. This theorem follows from the determinant
condition ${\rm det}(H-EI)=0$ on the eigenvalues and does not refer to the
solutions to the eigenvector equation $H\psi=E\psi$. Thus, $\cP\cT$-invariant
Hamiltonians that have Jordan-block structure still have a real secular
equation. Such secular equations could have complex-conjugate pairs of
solutions. However, such pairs would not be degenerate. Since Jordan-block
Hamiltonians cannot have nondegenerate eigenvalues, the eigenvalues of $\cP
\cT$-invariant Jordan-block Hamiltonians have no choice but to be real. This
establishes the reality of the energy eigenvalues of $\cP\cT$-symmetric
Jordan-block Hamiltonians in the Jordan-block sector(s).

As one varies the parameters in a Hamiltonian such as that in (\ref{E1}) to
transit from real to complex eigenvalues, one passes through a point where the
two energy eigenvalues become degenerate. Since there is such a transition at
that point, the energy spectrum has a square-root-branch-point singularity at
${\bf h}\cdot{\bf h}=0$; because of this singularity, the $\cC$ operator must
also be discontinuous and become undefined. At the transition two things can
occur: Either the two degenerate eigenvalues can have independent eigenvectors,
or the two eigenvectors can merge. By adjusting the parameters in (\ref{E1})
accordingly, one can achieve either of these two outcomes \cite{R8}. The special
points at which the transitions occur are thus points at which all energy
eigenvalues are real. At such points there is no need to find a criterion that
would force the eigenvalues of a $\cP\cT$-symmetric Hamiltonian to be real since
they already are. Thus, at such transition points the reality of the eigenvalues
of a $\cP\cT$-symmetric Hamiltonian is guaranteed. At points other than these
special points, the reality or complexity of eigenvalues is fixed by the
vanishing or nonvanishing of the $[\cC,\cP\cT]$ commutator \cite{R9}.

\section{The Generalized $\cQ$ Operator}
\label{S5}

In  $\cP\cT$ studies some special cases have been found (see \cite{R5} and
references therein) in which all energies are real and one can set $\cP^{-1}H\cP
=H^{\dag}$ and thus $\cP^{-1}\cC^{-1}H\cC\cP=H^{\dag}$ (because $[H,\cC]=0$). In
some of those cases the operator product $\cC\cP$ can be written as $\cC\cP=e^{
\cQ}$, where $\cQ$ is Dirac Hermitian \cite{R5a}. In these instances, the
similarity-transformed Hamiltonian $e^{-\cQ/2}He^{\cQ/2}$ is Dirac Hermitian.
However, no rule has been given that would determine the circumstances under
which these properties might hold or how they might generalize.

The energies of $\cP\cT$-symmetric Hamiltonians are real or appear in
complex-conjugate pairs. For such Hamiltonians $H$ and $H^{\dag}$ have the same
energy eigenspectra and must be related by a similarity transformation. The
requisite similarity transform generalizes the relation $\cP^{-1}H\cP=H^{\dag}$,
even when there are complex energies. To find the appropriate generalization, we
note that in the $2\times 2$ case the operator $A=(\sigma_1 h_1+\sigma_3h_3)/(
h_1^2+h_3)^{1/2}$ with inverse $A^{-1}=A$ performs transposition according to
$A^{-1}\mbox{\boldmath $\sigma$}\cdot{\bf h}A=\mbox{\boldmath $\sigma$}^T\cdot{
\bf h}$. With the $\cT$ operator having the form $\cT=K\sigma_2\mbox{\boldmath
$\sigma$}\cdot{\bf t}$, we introduce an operator $B^{-1}=A^*\mbox{\boldmath $
\sigma$}^*\cdot{\bf t}\sigma_2^*$ and get $B^{-1}\cT=KA$ and thus $B^{-1}\cT H
\cT^{-1}B=H^{\dag}$. For $2\times2$ $\cP\cT$-symmetric Hamiltonians, we obtain
$B^{-1}\cP^{-1}H\cP B=H^{\dag}$. Therefore, the requisite similarity transform
is given by $\cP B$.

To generalize this derivation to $N$-dimensional matrices, we must find an
$N$-dimensional transposition operator. To this end we introduce the $N^2$
generators $\lambda_i$ of $U(N)$ as written in the $N$-dimensional fundamental
representation and set $H=\sum_{n=1}^{N^2}\lambda_n h_n$. Of the $U(N)$
generators, $N(N+1)/2$ are symmetric and real while $N(N-1)/2$ are antisymmetric
and imaginary. The anticommutator of any symmetric $SU(N)$ generator with any
antisymmetric generator vanishes. Therefore, an operator of the form $A=(\sum_i
\lambda_ih_i)/J$ as summed over the symmetric $SU(N)$ $\lambda_i$ generators
only with $J$ being an appropriate normalization factor will effect $A^{-1}(
\sum_n\lambda_n h_n)A=\sum_n\lambda_n^T h_n$. This gives the transposition
operator.

To express the operator $\cC\cP$ in the form $e^{\cQ}$ with Dirac-Hermitian
$\cQ$, it is necessary that $\cC\cP$ be Hermitian and that it be a {\it positive
operator} (that is, all of its eigenvalues are positive). In the two-dimensional
case the determinants of both $\cC$ and $\cP$ are equal to $-1$. Hence, the
determinant of $\cC\cP$ is $+1$. Thus, if we could show that $\cC\cP$ is
Hermitian and that its trace is positive, we could show that it is a positive
operator. Let us evaluate
\begin{eqnarray}
\cC\cP-(\cC\cP)^{\dag}={\bf c}\cdot {\bf p}+i\mbox{\boldmath $\sigma$}\cdot{\bf
c}\times {\bf p}-{\bf c}^*\cdot{\bf p}+i\mbox{\boldmath $\sigma$}\cdot{\bf c}^*
\times{\bf p}=2i {\bf c_I}\cdot{\bf p}+2i\mbox{\boldmath $\sigma$}\cdot{\bf c_R}
\times {\bf p}.
\label{E19}
\end{eqnarray}
Since we can set ${\bf c}={\bf h}/X$ when $Y=0$, since ${\bf h_R}\cdot{\bf h_I}=
0$, and since ${\bf p}$ can be chosen so that ${\bf h_R}$ is parallel to ${\bf
p}$ when the secular equation is real, we see that $\cC\cP-(\cC\cP)^{\dag}$ is
zero when $Y=0$. If all energies are real, then $\cC\cP$ is Hermitian. (When $X=
0$, one can show that $\cC\cP+(\cC\cP)^{\dag}=0$, with all eigenvalues of $\cC
\cP$ being pure imaginary.) In the real-energy sector we find that ${\rm Tr}(
\cC\cP)=2{\bf c_R}\cdot{\bf p}+2i{\bf c_I}\cdot{\bf p}$ is always real and
positive. Thus, in the real-energy sector $\cC\cP$ is always a positive operator
\cite{R10}.

CMB is supported by a grant from the U.S.~Department of Energy. PDM wishes to
thank Dr.~S.-W.~Tsai for helpful discussions.

{}

\begin{thebibliography}{}
\bibitem{R11} Because ${\rm det}(\cC)=-1$, one eigenstate of $H$ has $\cC$
eigenvalue $\eta=1$ while the other has $\eta=-1$. Since the $\eta$ are real,
for any  energy eigenstate $|\alpha\rangle$ we obtain $(\cC\cP\cT\cC\mp\cP\cT)|
\alpha\rangle=\eta_{\alpha}(\cC\cP\cT|\alpha\rangle\mp\eta_{\alpha}\cP\cT|\alpha
\rangle)$. If $\cP\cT|\alpha\rangle=|\alpha\rangle$ then $(\cC\cP\cT\cC-\cP\cT)|
\alpha\rangle=0$, while if $\cP\cT|\alpha\rangle=|\beta\rangle$ where $|\beta
\rangle$ has $\eta_{\beta}=-\eta_{\alpha}$ we obtain $(\cC\cP\cT\cC+ \cP\cT)|
\alpha\rangle=0$. Thus, as noted in Sec.~\ref{S1}, if $\cP\cT|\alpha\rangle=|
\alpha\rangle$, for each energy eigenstate the energies are real; if $\cP\cT|
\alpha\rangle=|\beta\rangle$, the energies are complex.
\bibitem{R8} The only way a two-dimensional matrix with two equal eigenvalues
can avoid being Jordan block is the special case in which it has no nondiagonal
terms at all.
\bibitem{R9} Although it is beyond the scope of this paper, $\cP\cT$ symmetry
provides a natural framework for treating decays, with a continuous change in
parameters converting stable states into unstable ones. Thus, $\cP\cT$-symmetric
Hamiltonians may be of relevance to open quantum systems.
\bibitem{R10} Since $\cC\cP$ does not commute with $H$, one cannot introduce a
diagonal basis to try to generalize any two-dimensional $\cC\cP$ result to the
general $N$-dimensional case. Rather one must work with the full $U(N)$
operators given above. We do not pursue this question here.
\end{thebibliography}
\end{document}